# Atomic Structure of Benzene Which Accounts for Resonance Energy

(This work is dedicated to Kathleen Lonsdale)

[1] **Raji Heyrovska**

[1] *Institute of Biophysics, Academy of Sciences of the Czech Republic, Královopolská 135, 61265 Brno, Czech Republic. Email:* rheyrovs@hotmail.com

Receipt date:

ABSTRACT:

About eight decades ago, Lonsdale found from her crystallographic studies that benzene consists of carbon atoms of radii 0.71 Å (as in graphite) in an almost planar hexagon ring. On subtracting this radius from the known CC bond length (~ 1.39 Å) in benzene, one obtains the covalent double bond radius of carbon. This shows that the CC bond distance in benzene is simply the sum of the atomic radii of the adjacent two carbon atoms. Since this structure also explains the bond energy sum for benzene, the existing two resonating Kekule structures (not observed yet) are unnecessary.

1. INTRODUCTION

Benzene is an important aromatic compound, and yet its structure has been a mystery ever since its discovery [1a]. In an article commemorating Pauling, Stephen Mason [2a] writes: "Pauling's disciple, George Wheland, remarked that the benzene molecule is analogous to the real animal, the rhinoceros, described by a medieval traveler as a cross between two mythical beasts, the dragon and the unicorn". Commenting on the same subject, Servos [2b] writes: "Like the elephant in the tale of the blind men, the benzene molecule did not correspond to any of its many descriptions, but rather was captured by all of them." The facts that all the CC bond lengths are equal (1.39 +/- 0.01 Å) although the planar hexagonal benzene is supposed to have three double bonds of lengths 1.34 Å each alternating with and three single bonds of lengths 1.54 Å (as in diamond) each, and that the sum of the bond energies is less than the observed value, was explained as due to resonance [3] between the two Kekule structures [1b] shown in Fig. 1. See [1c] for some historical development in this field. In recent years, the author has found [4,5] that bonding distances in many chemical compounds including benzene are sums of the radii of the adjacent atoms or ions. This paper presents some existing experimental evidence in favor of the interpretation [4,5] of the atomic structure of benzene. The conclusion is that it was unfortunate not to have given more thought to Lonsdale's crystallographic data eight decades ago, which had the right answer to the structure of benzene. Further, the results show that it is no longer necessary to assume the two resonance structures for benzene, which have not (yet) been observed.

2. RECENT WORK ON THE ADDITIVITY OF ATOMIC RADII IN BENZENE

The recent finding [4] of the additivity of atomic covalent and or ionic radii in the inter-atomic and inter-ionic distances, led to the suggestion in [4,5] that the bond length, 1.39 Å in benzene is the sum of the radius, 0.72 Å (= $R_{r.b.}$, resonance bond radius, subscript: r.b.) of one carbon atom (with delocalized charge as in graphite [3] and graphene [1d, 6]) and the double bond (subscript: d.b.) radius ($R_{d.b.}$) 0.67 Å, [3] of the adjacent atom. Thus, it was shown [4,5] that benzene does not involve the CC single bond (subscript: s.b.) of the diamond type of radius ($R_{s.b.}$) 0.77 Å [3], which is found in aliphatic compounds like e.g., methane as in Fig. 2 (a).

The atomic structure of benzene [5] can be seen in Fig. 2 (b), where all the six bond lengths have the same value, $(R_{d.b.} + R_{r.b.}) = 0.67 + 0.71 = 1.38$ Å. Note that the empty space in the center of the hexagon fits an inscribed circle of radius 0.67 Å. The bond lengths in the aromatic rings in many biological compounds like the molecular components of nucleic acids [7], caffeine related compounds [8] and amino acids [9] confirm the structure of the benzene ring in Fig. 2 (b). In the following sections, some existing experimental evidences for the new structure of benzene have been provided.

3. SUPPORT FOR THE RESONANCE BOND AND DOUBLE BOND RADII

On looking through the literature for some support for the proposed atomic structure of benzene, it was found that Lonsdale [10] suggested a diameter close to that in graphite, 1.42 Å for each of the six carbon atoms, for a nearly planar hexagon in the aromatic ring in hexachlorobenzene. Cox [11] also used this value (0.71 Å) for the radius of carbon in benzene. Since the observed CC bond length [3] in benzene is 1.39 (+/- 0.01) Å, the remainder, 1.39 - 0.71 = 0.68 Å ~ 0.67 Å, the double bond radius of C, [3].

4. SUPPORT FOR TWO BONDS $C_{d.b.}$-H AND $C_{r.b.}$-H OF DIFFERENT LENGTHS

The structure of benzene in Fig. 2 (b) shows that there are three $C_{d.b.}$- H bonds of length 1.04 Å and three $C_{r.b.}$- H bonds of length 1.08 Å, differing (slightly) by 0.04 Å. Although the CH bond distances reported in the literature average around 1.08 Å, [3] and are with less accuracy than the CC bond distances, two distinct distances have been reported: 1.085 (+/- 0.017) Å and 1.059 (+/- 0.015) Å in [12] and 0.95 Å and 1.08 Å in [13]. Support is found also in [10] by the observation of two different distances for the carbon to chlorine distances of 1.67 and 1.79 Å. On subtracting the covalent radius of chlorine atom 0.99 Å [3] from the C-Cl distance, 1.67 Å, one gets 0.68 Å which is the double bond radius of carbon. The second C-Cl distance of 1.79 Å is close to the inter-ionic distance (see Table 3 in [4]), $d(C+) + d(Cl-) = 0.55 + 1.22 = 1.77$ Å, where $d(C+)$ and $d(Cl-)$ are the Golden ratio ($\varphi$) based ionic radii of C and Cl. Note from [4] that $d(C+) = 0.55 = d(CC)/\varphi^2 = 1.42/2.618$ Å, where $d(CC) = 1.42$ Å $= 2R_{r.b.}$ is the interatomic distance in graphite. Lonsdale [10] also mentions that the carbon atoms in benzene are probably polarized. [As a side note: $R_{r.b.}$ (= 0.71 Å) is related to $R_{s.b.}$ (= 0.77 Å) and $R_{d.b.}$ (= 0.67 Å) through the Golden ratio ($\varphi$) [4] as follows: $R_{r.b.} = (R_{s.b.}/\varphi^2) + (R_{d.b.}/\varphi) = (0.294 + 0.414)$ Å.]

5. SUPPORT FOR THE ATOMIC STRUCTURE: NO 'SURPLUS' BOND ENERGY

The bond energy sum for the benzene molecule in Fig. 1 assuming three C=C double bonds (146 kcal/mole [14]), three C-C single bonds (83 kcal/mole [14]) and six CH bonds (99 kcal/mole [14]) gives 1281 kcal/mole. This is less than the experimental value of 1323 kcal/mole [3] by 42 kcal/mole. This observed excess energy is attributed to resonance [3].

As per the structure in Fig. 2 (b), this surplus is of the right amount since the carbon atoms with resonance bonds (with delocalization of charge) as in graphene shown in Fig. 2 (c) have a higher energy (124.6 kcal/mole, [15]) than the single bonds as assumed in the calculations in [3]. On using the mean of the resonance bond energy and double bond energy $(146 + 124.6)/2 = 135.3$ kcal/mole for each of the six bonds, the difference, $1323 - 6 \times 135.3 = 511.2 = 6 \times 85.2$. Thus, the six CH bonds will each have an average energy of 85.2 kcal/mole. This agrees well with the bond dissociation energy (85 kcal/mole) for H-benzyl in [15]. Therefore, the author concludes that Fig. 2 (b) represents the most probable atomic structure for benzene. This conclusion could have been drawn soon after Lonsdale's work nearly eight decades ago. It is unfortunate that Pauling did not take serious notice of Lonsdale's work (although he treated covalent bond lengths as sums of atomic radii) and was publishing his new idea of resonating structures for benzene at about the same time [2b,16].

6. GRAPHENE, DIFFERENCE FROM BENZENE

Although graphite consists of two dimensional hexagonal array of carbon atoms called graphene [1d,6], the hexagons are structurally different from that of benzene. The structure of graphite is given in [3] with two double bonds and four single bonds within the ring and the same number connecting with six carbons of the outer rings. However, all the bond lengths are known to be equal [1d,6] to about 1.42 Å, despite the conventional notion that they have double bond and single bonds (like benzene which consists of one hexagon of carbon atoms with equal bond distances of 1.39 Å). In [5], the equality of the bond lengths in graphene was represented by the regular hexagon shown here in Fig. 2 (c) consisting of six atoms of equal resonance bond radii, 0.71 Å. The central inscribed empty circle is of the same radius (0.71 Å) as the six atoms of the hexagon, (whereas in benzene in Fig. 2 (b), the inscribed circle has a smaller radius, 0.67 Å that of the carbon atom with the double bond radius).

ACKNOWLEDGMENTS: The author thanks Prof. E. Palecek of the IBP, Academy of Sciences of the Czech Republic for the moral support and the IBP for financial support. Thanks are also due to Prof. A. K. Geim for the reference [6] and to Dr. T. Allgood of CCDC, Cambridge, England for promptly providing some crystallographic data pertaining to references [11 – 13] and more.

FIGURE CAPTIONS:

FIG. 1. The conventional resonance structure of benzene [1b, 3].

FIG. 2. Atomic structures of (a) methane, (b) benzene and (c) graphene [5]. (Subscripts, s.b.: single bond, d.b.: double bond and r.b.: resonance bond)

FIG. 1 (R.H.)

**Conventional structure of benzene**

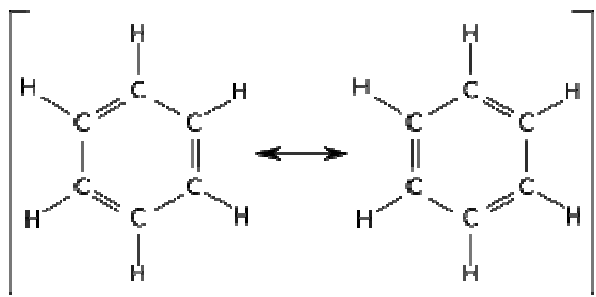

FIG. 2 (R.H.)

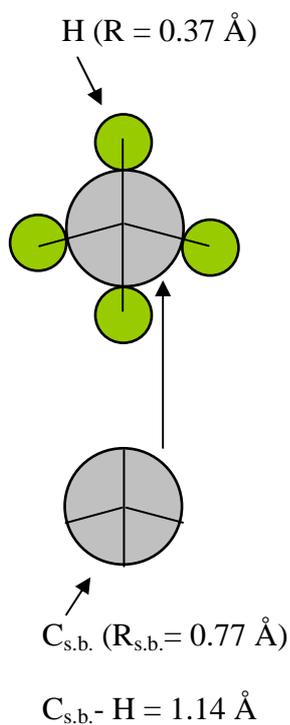
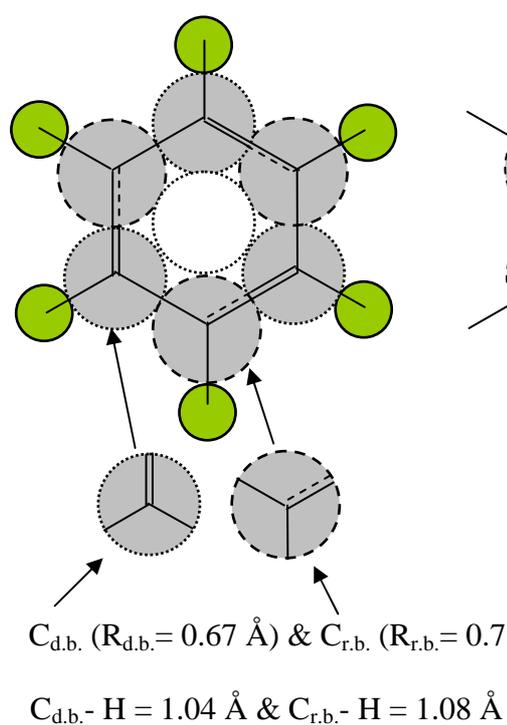
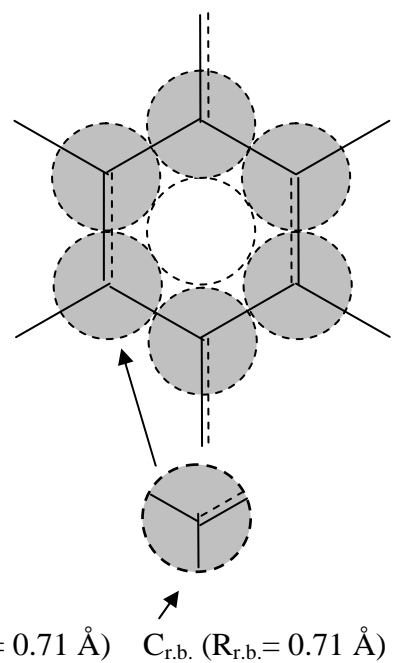

**(a) Methane**

H (R = 0.37 Å)

$C_{s.b.}$ ($R_{s.b.}$ = 0.77 Å)

$C_{s.b.}$- H = 1.14 Å

**(b) Benzene**

$C_{d.b.}$ ($R_{d.b.}$ = 0.67 Å) & $C_{r.b.}$ ($R_{r.b.}$ = 0.71 Å)

$C_{d.b.}$- H = 1.04 Å & $C_{r.b.}$- H = 1.08 Å

**(c) A graphene hexagon**

$C_{r.b.}$ ($R_{r.b.}$ = 0.71 Å)